\documentclass[fleqn,12pt,twoside]{article}
\usepackage{espcrc1}

\usepackage{epsf}
\usepackage[figuresright]{rotating}

\newcommand{\AmS}{{\protect\the\textfont2
  A\kern-.1667em\lower.5ex\hbox{M}\kern-.125emS}}

\title{Determining the BDMPS transport coefficient via medium-modified 
fragmentation functions}       

\author{\underline{Carlos A. Salgado}
 and Urs Achim Wiedemann\\[2ex]
Theory Division, CERN, CH-1211 Geneva 23, Switzerland.
                }
       
\begin{document}

\maketitle

\begin{abstract}
In nucleus-nucleus collisions at RHIC and LHC, partons produced
at high transverse momentum can undergo multiple scattering within
the collision region prior to fragmenting into hadrons. We have
studied the resulting medium-modified fragmentation function based
on a calculation of the BDMPS-Z medium-induced gluon radiation for
a dense, expanding medium of small finite extension. Here we explain 
how the BDMPS transport coefficient $\hat{q}$ which measures the 
energy density attained in the collision, can be extracted from the 
observed modification of high-$p_\perp$ hadroproduction. We also 
comment on the significant remaining uncertainties in extracting 
$\hat{q}$ from data.
\end{abstract}
\\

To leading order in perturbative QCD, high-$p_\perp$ hadroproduction in
proton--proton collisions is described by the factorization formula
%
\begin{equation}
{d\sigma^h\over dp_t^2 dy}= K(\sqrt{s})\int{dz\over z^2}\int dy_2
 \sum_{i,j}\, x_1f^A_i(x_1,Q^2)\, x_2f^B_j(x_2,Q^2)
\, \frac{d\sigma^{ij\to k}}{d\hat t}  D_{k\to h}(z,Q^2)\, .
\label{eq1}
\end{equation}
\noindent
Here $f^A_i$, $f^B_j$ are the proton parton distribution functions (PDF), 
$d\sigma^{ij\to k} / d \hat t$ is the partonic cross section and 
$D_{k\to h}(z,Q^2)$ describes the fragmentation of a parton $k$ into the
hadron $h$ carrying a fraction $z$ of the momentum. Eq. (\ref{eq1})
leads to a fair description of the shape of high-$p_\perp$
hadronic spectra while the normalisation has to be adjusted by 
an energy-dependent $K$-factor (see e.g. \cite{kariheli}). 
Also to NLO, the disagreement between theory and experiment lies
essentially in an albeit reduced normalisation factor~\cite{Aurenche:1999nz}.

For nucleus--nucleus collisions, additional corrections arise due to 
parton density effects. At sufficiently high $p_\perp$ where the
factorised form (\ref{eq1}) is expected to hold, two corrections are 
known: i) Density effects in the {\it initial state} can be parametrised 
by nuclear PDFs for which global fits are available \cite{EKRS}. 
At RHIC energies, however, this is only a small correction~\cite{kariheli}
\cite{ina}. 
ii) Density effects in the {\it final state} affect the fragmentation of 
the high $p_\perp$ parton produced in the hard collision. Several
studies indicate that the dominant final state correction is due
to medium-induced gluon radiation and the ensuing parton energy loss.
If the high $p_\perp$ parton looses with probability 
$P_E(\epsilon)$ a fraction $\epsilon$ of its energy while propagating
through the medium, then its medium-modified fragmentation function 
can be written as~\cite{inawang,W,nos}
\begin{eqnarray}
  D_{k\to h}^{(\rm med)}(z,Q^2) = \int_0^1 d\epsilon\, P_E(\epsilon)\,
  \frac{1}{1-\epsilon}\, D_{k\to h}(\frac{z}{1-\epsilon},Q^2)\, .
  \label{eq2}
\end{eqnarray}
\noindent
We have calculated the {\it quenching weights} $P_E(\epsilon)$ for
quarks and gluons starting from the BDMPS-Z 
\cite{BDMPS,Zakharov:1997uu,Baier:1998yf,urs} 
gluon radiation spectrum $dI/d\omega$ generalised to the case of an 
expanding medium of small finite size~\cite{nos}. 
For the case of 
a static medium, the resulting finite size corrections to the
BDMPS-expression of $dI/d\omega$ are shown in Fig. 1. They 
become important for gluon energies much smaller than the 
characteristic gluon frequency $\omega_c=\hat{q}\, L$. The
latter depends on the in-medium path-length $L$ and the BDMPS 
transport coefficient $\hat q$ which measures the amount of 
transverse momentum squared transferred from the medium to the 
hard parton per unit path-length. For a static medium, the
radiation spectrum depends only on $\omega_c$ and the dimensionless
parameter combination $R=\frac{1}{2} \hat{q}\, L^3$. As seen in
Fig. 1, the BDMPS case is the limit $L\to \infty$ of
our calculation  keeping $\omega_c$ finite. 
\begin{figure}[htb]
\begin{minipage}[t]{80mm}
\begin{center}
    \vskip -1.2cm
    \epsfxsize=7cm \leavevmode \epsffile{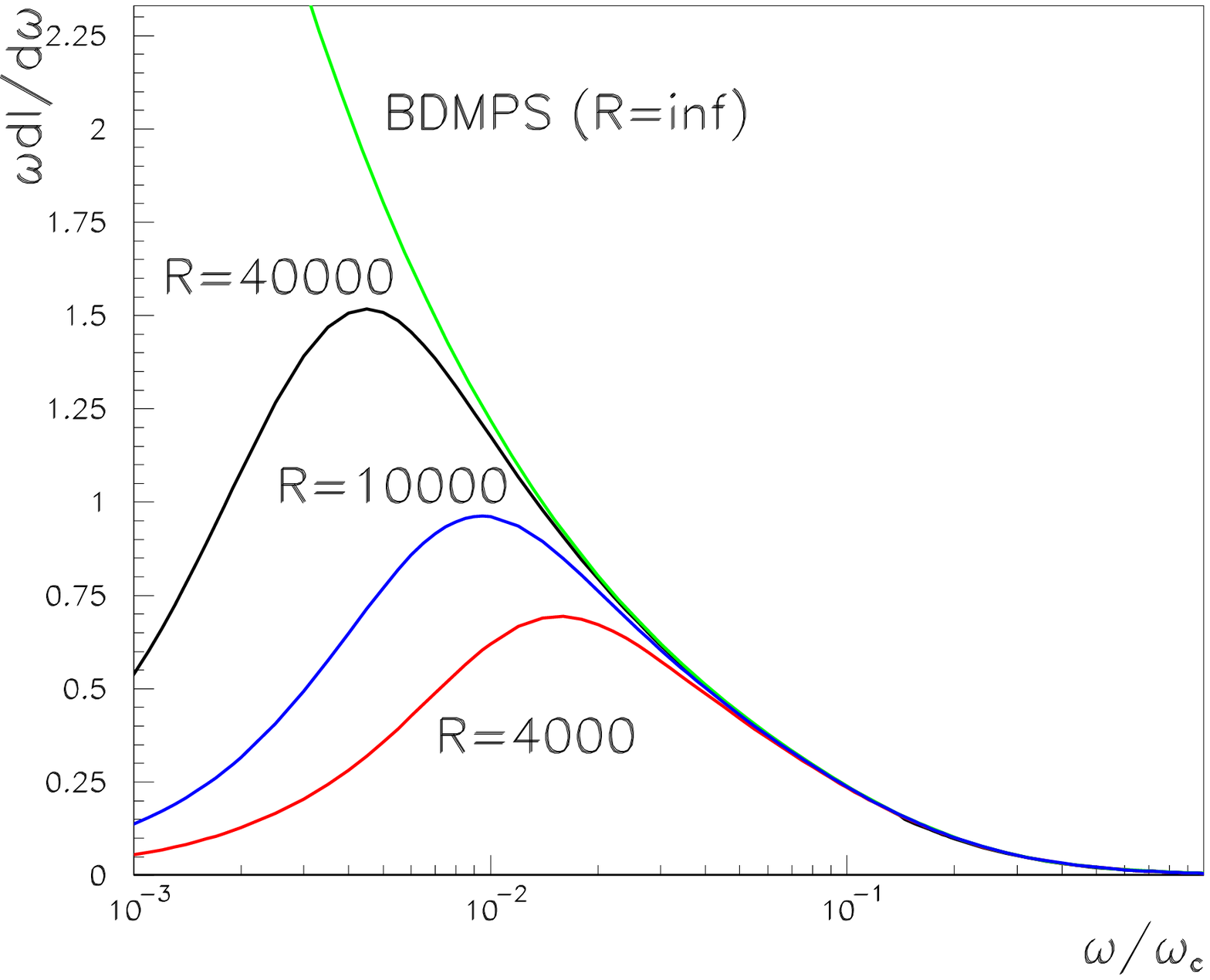}
  \end{center}
\vskip -1.5cm
\caption{(Up) Gluon radiation spectrum, $\omega dI/d\omega$, 
for several values of $R$ and comparison
with the BDMPS result.}
\label{fig:largenenough}
\vskip -0.5cm
\caption{(Right)
Continuous (upper figure) and discrete (lower)
contributions to the quenching weight $P_E(\epsilon)$
of a fast quark for different values of $R$.}
\label{fig:toosmall}
\end{minipage}
\hspace{\fill}
\begin{minipage}[t]{75mm}
\begin{center}
    \vskip -1.7cm
    \epsfxsize=7cm \leavevmode \epsffile{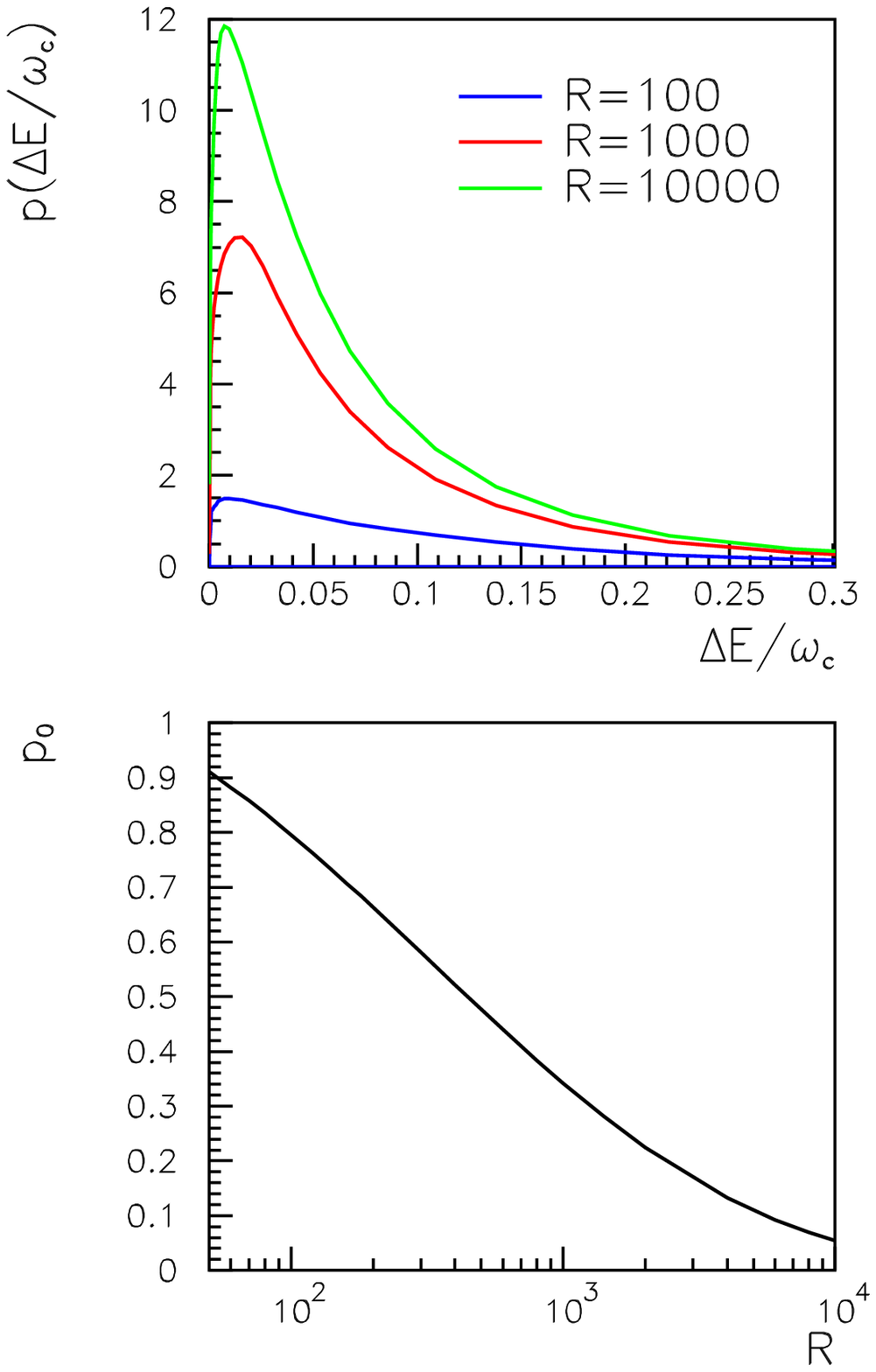}
  \end{center}
\end{minipage}
\end{figure}

Quenching weights are evaluated in the independent gluon
emission approximation \cite{lastBDMS} and show both a 
discrete and a continuous contribution (see Fig. 2),
\begin{eqnarray}
  P_E(\epsilon) &=& \sum_{n=0}^\infty \frac{1}{n!}
  \left[ \prod_{i=1}^n \int d\omega_i \frac{dI(\omega_i)}{d\omega}
    \right]
    \delta\left(\epsilon -\sum_{i=1}^n {\omega_i \over E} \right)
    \exp\left[ - \int d\omega \frac{dI}{d\omega}\right]
\label{eq3}\\
  &=& p_0\delta(\epsilon)+p(\epsilon)\, .
  \label{eq4}
\end{eqnarray}
Here, $P(\epsilon)$ is a generalised probability of norm 1. For
a medium of finite length $L$, there is a finite probability $p_0$ 
that the fragmentation is not disturbed by the medium. This
discrete contribution decreases with increasing in medium path-length
or increasing density of the medium.

The transport coefficient is proportional to the density of
scattering centres, $\hat{q}(\tau) \propto n(\tau)$. The dynamical
expansion of the collision region can be parametrised by a power law
decrease in proper time, $n(\tau)=n_0\left(\frac{\tau_0}{\tau}\right)^\alpha$,
where $\alpha=0$ for static, $\alpha=1$ for Bjorken expansion. 
Fig. 3 shows the gluon radiation spectrum for different values 
of $\alpha$ with parameters $\omega_c$ and $R$ written in terms 
of the linearly weighted transport coefficient
\begin{eqnarray}
\overline{\hat{q}}=\frac{2}{L^2}\int_{\tau_0}^{\tau_0+L} d\tau\ (\tau-\tau_0)\
\hat{q}(\tau)\, .
\label{eq6}
\end{eqnarray}
After rescaling according to (\ref{eq6}), the radiation spectra 
$dI/d\omega$ calculated for different values $\alpha$ coincide (see Fig. 3).
This establishes a dynamical scaling law~\cite{nos} which simplifies
calculations considerably since results for a dynamically expanding medium
can be obtained by scaling the result obtained for a static medium. It
extends earlier findings of dynamical scaling of the average energy 
loss~\cite{Baier:1998yf,W,Wang:2002ri}
to the $\omega$-differential radiation spectrum~\cite{nos}.

\begin{figure}[htb]
\begin{minipage}[t]{80mm}
    \vskip -1.2cm
\begin{center}
   \epsfxsize=6.5cm \leavevmode \epsffile{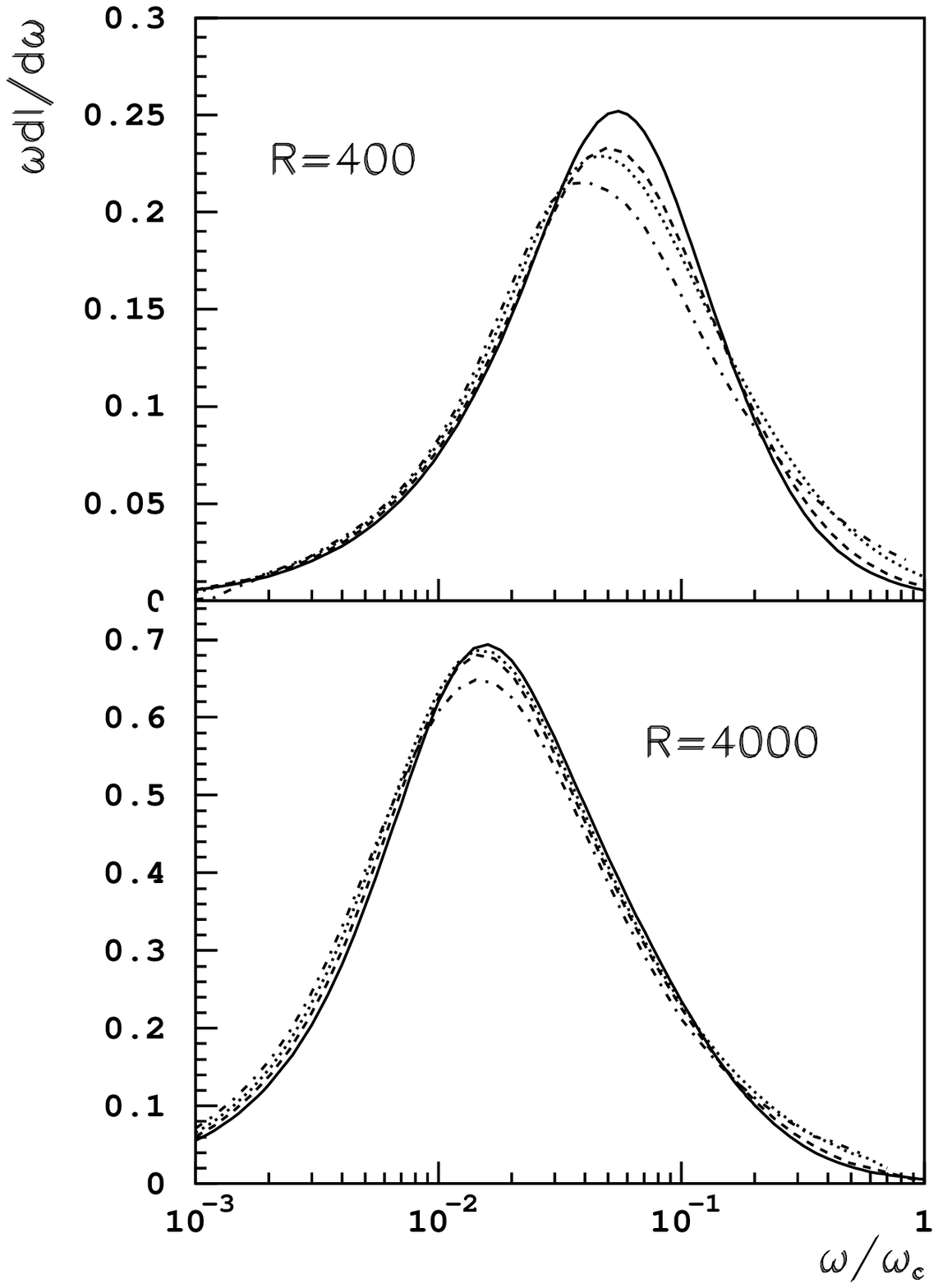}
\end{center}
\vskip -1.3cm
\caption{The gluon radiation spectrum, $\omega dI/d\omega$,
plotted as a function of $\omega_c = \overline{\hat{q}} L$
and $R = \frac{1}{2} \overline{\hat{q}} L^3$. Different 
curves correspond to different expansion parameters
$\alpha = 0$, $0.5$, $1.0$ and $1.5$.}
\label{fig3}
\vskip -0.5cm
\end{minipage}
\hspace{\fill}
\begin{minipage}[t]{75mm}
    \vskip -1.3cm\hskip -0.7cm
    \epsfxsize=9cm \leavevmode \epsffile{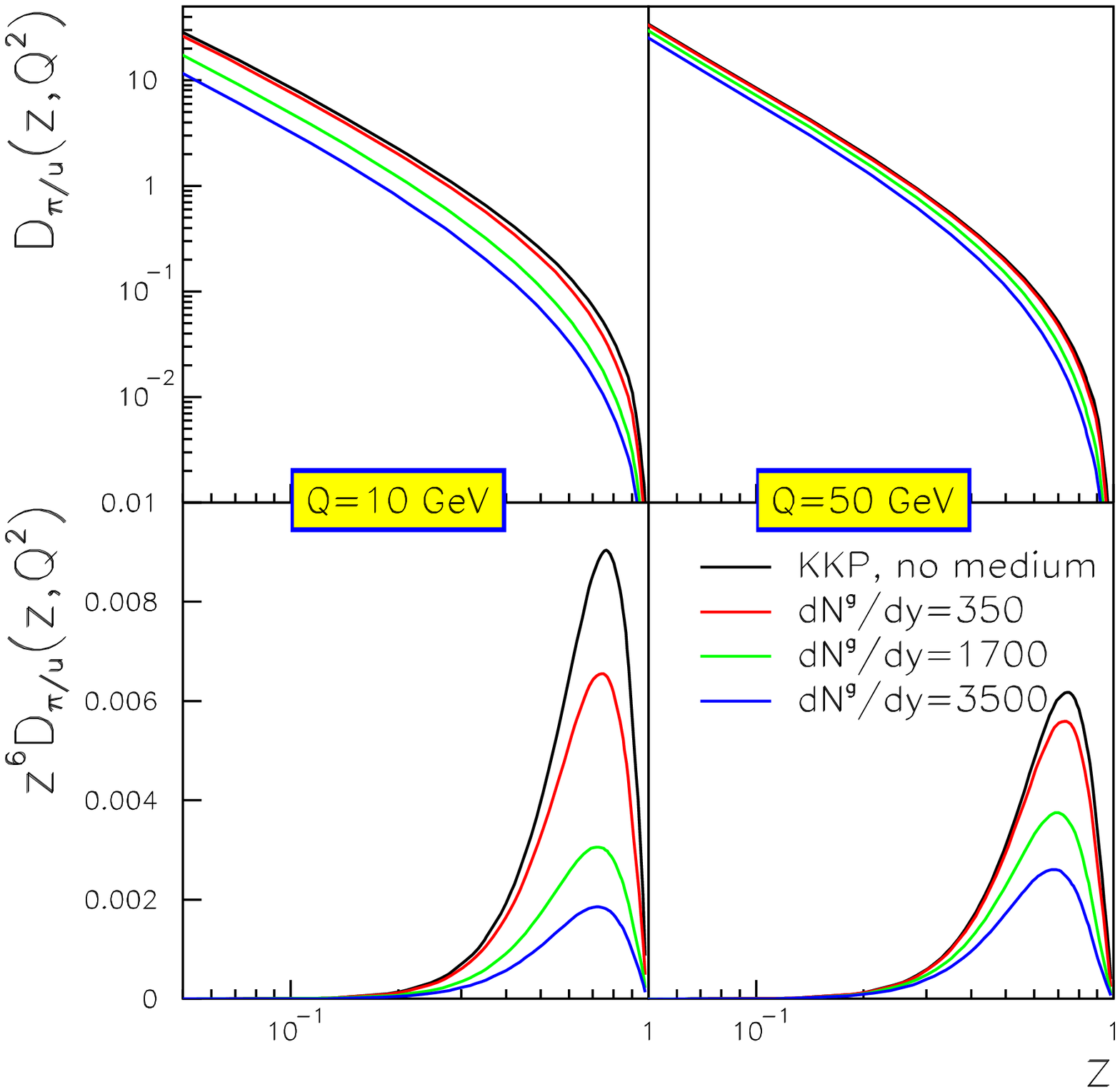}
\vskip -1.3cm
\caption{
The LO KKP~\cite{KKP} fragmentation function $u\to\pi$ for no medium
and the medium-modified fragmentation functions for different
gluon rapidity densities in a medium of length $L=7$ fm.
}
\label{fig4}
\end{minipage}
\end{figure}
Based on the quenching weights (\ref{eq3}) for static and 
expanding media, we have calculated medium-modified fragmentation 
functions according to (\ref{eq2}), using the KKP 
parametrisation~\cite{KKP} for parton fragmentation in the vacuum
(see Fig.~\ref{fig4}). The virtuality $Q$ of the parent parton 
is identified with its initial transverse energy. For a medium
showing one-dimensional Bjorken expansion ($\alpha = 1$), the
transport coefficient $\overline{\hat{q}}$ is related to the
gluon rapidity density as~\cite{W,BDMPS,nos}
\begin{eqnarray}
R=\frac{1}{2}\overline{\hat{q}}L^3=\frac{L^2}{R_A^2}\frac{dN^g}{dy}\, .
\label{eq7}
\end{eqnarray}
The calculation presented in Fig.~\ref{fig4} allows to determine
medium modifications to the transverse momentum spectrum
(\ref{eq1}) at sufficiently high $p_\perp$. While this last step
is not yet performed, one can estimate the effect based on the
observation~\cite{kariheli} that the hard partonic cross section 
in (\ref{eq1}) essentially weighs the fragmentation function with 
the power $z^6$. Inspecting the corresponding unintegrated moment
$z^6 D_{\pi/u}(z,Q^2\sim p_\perp^2)$ in Fig.~\ref{fig4}, one estimates
that a factor 2 suppression of the hadronic pion spectrum for 
$p_\perp\sim$ 6--7 GeV corresponds to an initial gluon density of 
500--1000. For an in-medium path-length of $L = 7$ fm, this 
amounts to a line-averaged transport coefficient $\overline{\hat{q}} = $ 
(300 MeV)$^2$ -- (600 MeV)$^2$/fm which corresponds to 
$\langle p_\perp^2 \rangle / L_{\rm med} $ of several GeV$^2$/fm within the
first two fm/c.  

For a reliable determination of $\hat{q}$ from data, the evaluation
of (\ref{eq1}) is not sufficient. One also has to discriminate the
medium modification discussed here from other corrections to (\ref{eq1}), 
such as ``soft'' hadronic contributions at low $p_\perp$ and higher 
order perturbative contributions at high $p_\perp$. Much of this 
can be done by studying the dependence of spectra on the in-medium 
path-length e.g. via the centrality dependence of high-$p_\perp$
hadroproduction and monojet/dijet production. Another discriminatory
tool results from the fact that medium-induced gluon radiation 
is a medium-enhanced power correction which shows a power law
decrease with increasing $p_\perp$. This is in contrast e.g. to
perturbative NLO corrections which mainly affect the normalisation
but not the shape of spectra. This will also provide an important
additional cross check at LHC: according to (\ref{eq7}), if the rapidity 
density increases by a factor 5 from RHIC to LHC, and if the
suppression factor 2 for RHIC $\pi^0$ spectra~\cite{David:2001gk}
at $p_\perp \sim $ 6--7 GeV
can be attributed solely to medium-induced gluon radiation, then
a factor 2 suppression of the corresponding spectrum at LHC is
expected in the range $p_\perp \sim $ 40--50 GeV. At even
higher $p_\perp$ still accessible at LHC, this suppression would
decrease according to a power law, thus leading to a modified $p_\perp$-shape of hadronic
spectra.

C.A.S. is supported by a Marie Curie Fellowship of the European
Community programme TMR (Training and Mobility of Researchers), under
the contract number HPMF-CT-2000-01025.

\end{document}